\documentclass[12pt]{article}
\usepackage{tikz}
\usepackage{pgf}
\usepackage[colorlinks=true,linkcolor=black, citecolor=black,urlcolor=black]{hyperref}

\usepackage{multirow,graphics}
\usepackage{enumerate}
\usepackage{amstext}
\usepackage{amssymb}
\usepackage{amsmath}
\usepackage{booktabs}
\usepackage{graphicx}
\usepackage{subcaption}
\usepackage{color}
\usepackage[nodisplayskipstretch]{setspace}
\usepackage{tcolorbox}
\usepackage{float}
\usepackage{wasysym}
\usepackage{ytableau}
\usepackage{textcomp}

\usetikzlibrary{decorations.pathmorphing}
\usetikzlibrary{decorations.markings}
\usetikzlibrary{intersections}
\usetikzlibrary{calc}
  \usetikzlibrary{positioning}
\usetikzlibrary{shapes}
\usetikzlibrary{fit}
\usetikzlibrary{backgrounds}

\tikzset{
photon/.style={decorate, decoration={snake}},
particle/.style={postaction={decorate},
    decoration={markings,mark=at position .5 with {\arrow{>}}}},
antiparticle/.style={postaction={decorate},
    decoration={markings,mark=at position .5 with {\arrow{<}}}},
gluon/.style={decorate, decoration={coil,amplitude=2pt, segment length=4pt},color=purple},
wilson/.style={color=blue, thick},
scalarZ/.style={postaction={decorate},decoration={markings, mark=at position .5 with{\arrow[scale=1]{stealth}}}},
scalarX/.style={postaction={decorate}, dashed, dash pattern = on 4pt off 2pt, dash phase = 2pt, decoration={markings, mark=at position .53 with{\arrow[scale=1]{stealth}}}},
scalarZw/.style={postaction={decorate},decoration={markings, mark=at position .75 with{\arrow[scale=1]{stealth}}}},
scalarXw/.style={postaction={decorate}, dashed, dash pattern = on 4pt off 2pt, dash phase = 2pt, decoration={markings, mark=at position .60 with{\arrow[scale=1]{stealth}}}},
frozen/.style={inner sep=0.7mm, rectangle,draw},
frozenblue/.style={rectangle, draw, fill=blue!20, inner sep=0.7mm},
blue/.style={rectangle, rounded corners, fill=blue!20, inner sep=0.7mm},
red/.style={rectangle, rounded corners, fill=red!20, inner sep=0.7mm},
>=stealth,
norm/.style={->, draw, shorten <=2pt, shorten >=2pt},
diag/.style={->, draw, shorten <=5pt, shorten >=3pt},
every node/.style={inner sep=0.5mm}
}

\newcommand\hns[1]{\text{hns}[#1]}

\newcommand\hnsqbar[1]{\text{hns}_{\bar Q}[#1]}

\newcommand\be{\begin{equation}}
\newcommand\ee{\end{equation}}

\newcommand{\la}{\langle}
\newcommand{\ra}{\rangle}
\newcommand{\ab}[1]{\la #1 \ra}

\newcommand{\cB}{\begin{cal}B\end{cal}}
\newcommand{\cC}{\begin{cal}C\end{cal}}

\newcommand{\cE}{\begin{cal}E\end{cal}}

\newcommand{\cN}{\begin{cal}N\end{cal}}
\newcommand{\cO}{\begin{cal}O\end{cal}}

\newcommand{\cR}{\begin{cal}R\end{cal}}
\newcommand{\cS}{\begin{cal}S\end{cal}}

\newcommand{\as}[2]{
  $a_{#1#2}$
}

\newcommand{\bdiamond}[1][fill=black]{\tikz [x=1.2ex,y=1.2ex,line width=.1ex,line join=round, yshift=-0.285ex] \draw  [#1]  (0,0) -- (.5,.5) -- (0,1) -- (-0.5,0.5)  -- cycle;}%
\newcommand{\wdiamond}[1][fill=white]{\tikz [x=1.2ex,y=1.2ex,line width=.1ex,line join=round, yshift=-0.285ex] \draw  [#1]  (0,0) -- (.5,.5) -- (0,1) -- (-0.5,0.5)  -- cycle;}%
\newcommand{\bcircle}[1][fill=black]{\tikz [x=1.2ex,y=1.2ex,line width=.1ex,line join=round, yshift=-0.285ex] \draw  [#1]  (0,0) circle (0.5);}%
\newcommand{\wcircle}[1][fill=white]{\tikz [x=1.2ex,y=1.2ex,line width=.1ex,line join=round, yshift=-0.285ex] \draw  [#1]  (0,0) circle (0.5);}%

\normalsize
\makeatletter
\divide\@tempdima\baselineskip
\@tempcnta=\@tempdima
\skip\@mpfootins = \skip\footins
\let\svthefootnote\thefootnote
\newcommand\blankfootnote[1]{%
  \let\thefootnote\relax\footnotetext{#1}%
  \let\thefootnote\svthefootnote%
}
\let\svfootnote\footnote
\renewcommand\footnote[2][?]{%
  \if\relax#1\relax%
    \blankfootnote{#2}%
  \else%
    \if?#1\svfootnote{#2}\else\svfootnote[#1]{#2}\fi%
  \fi
}

\renewcommand{\@dotsep}{10000}
\makeatother

\begin{document}
\null\vskip-43pt \hfill
\begin{minipage}[t]{30mm}
	DESY 18-214 
\end{minipage}
\numberwithin{equation}{section}
\begin{center}
\phantom{vv}

\vspace{3cm}
\bigskip

{\Large \bf    Cluster adjacency and the four-loop NMHV heptagon}

\bigskip
\mbox{{\bf James Drummond}$^1$, \bf{Jack Foster}$^1$,  \bf{\"Omer G\"urdo\u gan}$^1$,
  {\bf Georgios Papathanasiou}$^2$
}
\footnote[]{ 
 {\sffamily 
 \{\tt j.a.foster, j.m.drummond, o.c.gurdogan\}@soton.ac.uk }
}
\footnote[]{ 
 {\sffamily 
 \tt georgios.papathanasiou@desy.de 
 }
 }
\bigskip

{\em $^1$ School of Physics \& Astronomy, University of Southampton,\\
  Highfield, Southampton, SO17 1BJ, United Kingdom.}\\
\vskip .2truecm 
\vskip .2truecm
{\em $^2$ DESY Theory Group, DESY Hamburg,\\
  Notkestra\ss e 85, D-22607 Hamburg, Germany.}

\vspace{3cm}
\bigskip
\vspace{30pt} {\bf Abstract}
\end{center}

\noindent 
We exploit the recently described property of cluster adjacency for scattering amplitudes in planar $\mathcal{N}=4$ super Yang-Mills theory to construct the symbol of the four-loop NMHV heptagon amplitude. We use a manifestly cluster adjacent ansatz  
and describe how the parameters of this ansatz are determined using simple physical consistency requirements.
We then specialise our answer for the amplitude to the multi-Regge limit, finding agreement with previously available results up to the next-to-leading logarithm, and obtaining new predictions up to (next-to)$^3$-leading-logarithmic accuracy.

\newpage
\tableofcontents

\section{Introduction}
The idea of constructing scattering amplitudes from their analytic structure is a very old one. The idea developed in many directions, including the idea of the unitarity cut construction for loop amplitudes \cite{Bern:1994zx} and the BCFW recursion for tree-level amplitudes \cite{Britto:2005fq}. Related ideas have been applied to constructing the S-matrix of massive theories directly \cite{Paulos:2017fhb,Guerrieri:2018uew}, inspired by recent developments in the numerical bootstrap for conformal field theories \cite{Rattazzi:2008pe,Paulos:2016fap}.

Here we would like to develop further a theme that has seen a lot of progress in recent years, namely the perturbative bootstrap programme applied to the massless amplitudes of planar $\mathcal{N}=4$ super Yang-Mills theory \cite{Dixon:2011pw,Dixon:2011nj,Dixon:2013eka,Dixon:2014voa,Dixon:2014iba,Drummond:2014ffa,Dixon:2015iva,Caron-Huot:2016owq}. In particular we will apply the recently discovered property of cluster adjacency \cite{Drummond:2017ssj,Drummond:2018dfd} which exploits and develops the link between the singularities of the amplitudes and cluster algebras of a certain type \cite{Golden:2013xva}. 

The link opened up in \cite{Golden:2013xva} relates the branch-cut singularities of polylogarithmic functions (also known as the symbol `letters') with the $\mathcal{A}$-coordinates of cluster algebras associated to the kinematical configuration space ${\rm Conf}_n(\mathbb{P}^3)$. Cluster adjacency then dictates how the singularities are related to each other. In particular two symbol letters (or successive discontinuities) can only be present if the associated $\mathcal{A}$-coordinates appear together in some cluster. Such a relation was found by studying the known hexagon and heptagon loop amplitudes, including the four-loop MHV heptagon constructed in \cite{Dixon:2016nkn}.

As an example of the power and utility of the cluster adjacency principle we show here how it can be used to construct the four-loop NMHV heptagon amplitude from a rather minimal and manifestly cluster adjacent ansatz. We begin here with a very brief review of some basic aspects of scattering amplitudes in planar $\mathcal{N}=4$ super Yang-Mills theory. For the aspects not reviewed in this paper, we refer the reader to previous papers relevant to the cluster bootstrap programme where many aspects of amplitudes, bootstraps and cluster algebras have been discussed in great detail.

Scattering amplitudes in planar ${\cal N}=4$ super Yang-Mills of $n$
particles with momenta $\{p_i\}$ are dual to expectation values of (super) Wilson
loops on polygonal light-like contours \cite{Alday:2007hr,Drummond:2007aua,Brandhuber:2007yx,Drummond:2007cf,Drummond:2007bm,Bern:2008ap,Drummond:2008aq,Mason:2010yk,CaronHuot:2010ek} with the vertices satisfying
\begin{equation}
  p_i = x_{i+1} - x_i\,.
\end{equation}
This duality has very profound consequences for the scattering
amplitudes. In particular, the scattering amplitude exhibits the
anomalous conformal symmetry acting on the Wilson loop. The
scattering (super)amplitude then can be decomposed into two parts
\begin{align}\label{eq:BDS-like_normalisation}
  {\cal A}_n
  &=
  {\cal A}_n^{\text{BDS-like}}\,
  {\cal E}_n\, ,
  \end{align}
where ${\cal A}_n^{\text{BDS-like}}$ is the IR divergent BDS-like MHV superamplitude \cite{Alday:2009dv}.  It is the unique
solution to the anomalous dual-conformal Ward identity \cite{Drummond:2007au} dependent on only the two-particle invariants $(p_i + p_{i+1})^2 = x_{i\,i+2}^2$ (for $n\neq0$ mod. $4$). The remaining finite piece $ {\cal E}_n$ can be expanded into sectors,
    \begin{align}\label{eq:BDS-like_superamplitude}
  {\cal E}_n &= \mathcal{E}_{n,{\rm MHV}} + \mathcal{E}_{n,{\rm NMHV}} + \ldots 
\end{align}
and is invariant under the dual conformal symmetry. 

The function $\mathcal{E}_{n,{\rm MHV}}$ depends only on the cross-ratios of the Wilson loop,
\begin{equation}
  \label{eq:xratios}
  u_{ij}
  =
  \frac{x_{i j+1}^2 x_{j i+1}^2}
  {x_{ij}^2x_{i+1,j+1}^2}\, .
\end{equation}
For a seven-particle process, there are seven such cross-ratios
satisfying a Gram determinant constraint which makes ${\cal E}_{7, {\rm MHV}}$ a
six-variable function.  For NMHV amplitudes the function $\mathcal{E}_{n,{\rm NMHV}}$ is a sum over certain dual superconformal invariants $[ijklm]$ with coefficients which are functions of the dual conformal cross-ratios above. We will detail this structure further after reviewing cluster adjacent polylogarithms in Sect. \ref{sec:symb-alph-clust}. Our ansatz and computation for the four-loop NMHV heptagon amplitude is detailed in Sect \ref{sec-Comp}.

Having obtained our four-loop result we then analyse the amplitude in multi-Regge kinematics in Sect. \ref{sec-MRK}. The multi-Regge or high-energy limit is the arena where realistically occurring scattering configurations, originally studied within the analytic S-matrix programme \cite{White:2000zs} and QCD \cite{Kuraev:1976ge,Kuraev:1977fs,Balitsky:1978ic}, meet a beautiful simplification of their dynamical description in terms of effective, two-dimensional degrees of freedom. Particularly for $\cN=4$ SYM theory, which is the focus of this article, this simplicity has allowed for the identification of the space of functions required to describe the amplitude of an arbitrary number of external gluons $n$ at any loop order in the limit \cite{Dixon:2012yy,DelDuca:2016lad}, and in fact for $n=6$ has even led to the determination of the amplitude at finite coupling \cite{Basso:2014pla}. 

In this limit amplitudes develop large logarithms in some of the kinematic variables \eqref{eq:xratios}, and so at each loop order they reduce to a polynomial of these logarithms, the highest order of which corresponds to the leading logarithmic approximation or LLA, with an obvious generalisation to the (next-to)$^k$-leading logarithmic approximation or N$^k$LLA. The analysis of our four-loop amplitude provides a check of the consistency of our result with the expected structure of the Fourier-Mellin representation described in \cite{DelDuca:2016lad,DelDuca:2018hrv} at LLA and NLLA. It then also provides new predictions at the next two logarithmic orders. The four-loop results are provided in an ancillary file as are the new predictions for the amplitudes in multi-Regge kinematics.

\section{Symbol alphabet and cluster adjacency}
\label{sec:symb-alph-clust}

The amplitudes we investigate here are believed to be given in terms of multiple polylogarithms. These are iterated integrals with a specified set of singularities or `letters'. In our case the letters are described in terms of cluster $\mathcal{A}$-coordinates for cluster algebras of a particular type. Here we make use of the symbol, an algebraic object which captures the analytic structure of a polylogarithmic function. The entries of the symbol are drawn from the set of letters defining a given class of polylogarithms. We refer the reader to the many available references, e.g. \cite{Chen:1977oja,Goncharov:2005sla,Brown:2009qja,Goncharov:2010jf,Duhr:2012fh,Drummond:2018dfd} for background on polylogarithms and symbols.

In this section we summarise the statement of cluster adjacency and
its consequences for the symbols of amplitudes. We also review R-invariants and the
way they are related to the final entries of the symbols of NMHV amplitudes. Here we
will restrict the discussion to the facts relevant to the calculation
of heptagon amplitudes but we refer the reader to
\cite{Golden:2013xva} for a more in-depth exposure on cluster algebras
and to \cite{Drummond:2017ssj,Drummond:2018dfd} for cluster adjacency.

The kinematics of the scattering amplitudes or light-like Wilson-loops
are naturally parametrised using $n$ momentum twistors \cite{Hodges:2009hk}
$Z_i \in \mathbb{CP}_3$ with $i=1,\dotsc,n$. Each twistor also carries an index $A$ which indicates the linear action of the $sl_4$ dual conformal symmetry. The basic $sl_4$ invariants are the Pl\"ucker coordinates $\ab{ijkl}$ where
\begin{equation}
  \ab{ijkl}
  =
  \epsilon_{ABCD}
  Z_i^A Z_j^B Z_k^C Z_l^D\,.
\end{equation}
In special cases where the four labels of the Pl\"ucker coordinate
consists of two adjacent pairs, $\ab{i-1\, i \, j-1\, j}$, they correspond to multi-particle Mandelstam invariants
\begin{equation}
  \frac{
    \ab{i-1 \,i \,j-1\, j}
  }{
    \ab{i-1 \, i I}\ab{j-1 \,j I}
    }
  =
  x_{ij}^2\, ,
\end{equation}
where $I$ is the `infinity twistor' which is necessary to
relate twistor brackets to non conformally invariant quantities such
as Mandelstam invariants. However, a dual-conformal quantity such as
$\mathcal{E}_n$ depends only on the homogeneous rational combinations of
these brackets such that the dependence on the infinity twistor cancels
out.

The four-brackets $\langle ijkl \rangle$ are Pl\"ucker coordinates because the $n$ twistors $Z_i$ can be thought of as parameterising a Grassmannian ${\rm Gr}(4,n)$ modulo the rescaling of each of the $Z_i$ individually. Since the global rescaling of all $Z_i$ simultaneously is already taken into account in the definition of the Grassmannian, the kinematical space is identified with
\be
{\rm Conf}_n(\mathbb{P}^3) = {\rm Gr}(4,n)/(\mathbb{C}^*)^{n-1}\,.
\ee

The structure of the cluster algebras related to ${\rm Conf}_n(\mathbb{P}^3)$ dictates that the Pl\"ucker coordinates are the only multiplicatively
independent invariants for six particles, while for seven particles one finds that the symbol of
$\mathcal{E}_7$ contains also quadratic combinations such as
\begin{equation}
  \ab{i(jk)(lm)(np)}
  =
  \ab{ijkn}\ab{ilmp}
  -
  \ab{ijkp}\ab{ilmn}\,.
\end{equation}
Starting from eight particles, besides polynomials in Pl\"ucker
coordinates, also algebraic roots show up in the symbol of the
amplitude. These appear already at one loop due to the the presence of four-mass-box type cuts.

\subsection{Cluster-adjacent symbols}

The symbol entries encode the branch points of scattering amplitudes
and their analytic continuations across branch cuts. In an
$n$-particle process with dual-conformal invariance,
${\rm Conf}_n(\mathbb{P}^3)$ cluster algebras prescribe where in the
kinematical space these branch points are. In the case of
seven-particle scattering, the ${\rm Conf}_7(\mathbb{P}^3)$ cluster
algebra generates 49 branch points which appear in 42 homogeneous
combinations in a scattering amplitude. These 42 can be split into six
groups of seven which are closed under the $\mathbb{Z}_7$ cyclic
symmetry \cite{Drummond:2014ffa},
\begin{equation}
\begin{aligned}[b]
  a_{11} &= \frac{\langle 1234\rangle\langle1567\rangle\langle2367\rangle}{\langle1237\rangle\langle1267\rangle\langle3456\rangle}\\
  a_{31} &= \frac{\langle1567\rangle\langle2347\rangle}{\langle1237\rangle\langle4567\rangle}\\
  a_{51} &= \frac{\langle1(23)(45)(67)\rangle}{\langle1234\rangle\langle1567\rangle}
\end{aligned}
\,\,\,
\begin{aligned}[b]
  a_{21} &= \frac{\langle1234\rangle\langle2567\rangle}{\langle1267\rangle\langle2345\rangle}\\
  a_{41} &= \frac{\langle2457\rangle\langle3456\rangle}{\langle2345\rangle\langle4567\rangle}  \\
  a_{61} &= \frac{\langle1(34)(56)(72)\rangle}{\langle1234\rangle\langle1567\rangle}\,,
\end{aligned}\,
\label{heptletters}
\end{equation}
where the cyclic copies are defined as
$a_{i,j+r} = a_{ij}\bigr|_{Z_k \mapsto Z_{k+r}}$. The adjacent
brackets of the form $\langle i\, i+1\,i+2\,i+3 \rangle$ correspond to
frozen nodes that appear in every cluster and the factors of each of
the $a_{ij}$ include only a single active node.

The $a_{ij}$ above each correspond to a boundary component of the `positive' region in kinematical space. In the interior of this region all the $a_{ij}$ are positive. The positive region is also referred to as the cluster polytope (in the heptagon case it is the $E_6$ polytope). In the symbols of heptagon amplitudes, only the $a_{1i}$ may appear in the initial entries. This corresponds to imposing physical branch cuts on the Euclidean sheet.

To see all these coordinates in the context of the
${\rm Conf}_7(\mathbb{P}^3)$ cluster algebra, is it instructive to
consider one of the clusters with the topology of an $E_6$ Dynkin
diagram, which can be obtained after a series of mutations of the
initial cluster and contains nodes of all six types listed above.
  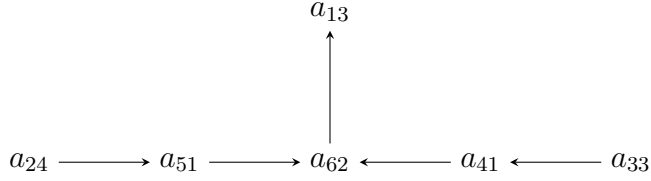
\begin{figure}
    \centering
      \begin{tikzpicture}
    [    unfrozen/.style={},
    frozen/.style={inner sep=1.2mm,outer sep=0mm,yshift=0},
    node distance = 0.4cm, scale =1, transform shape, remember picture
    ]
    
    \node[] (a51) at (-2, 0) {$a_{51}$}; 
    \node[] (a24) at (-4, 0)   {$a_{24}$};           
    \node[] (a62) at (0, 0)    {$a_{62}$};          
    \node[] (a41) at (2, 0)  {$a_{41}$};          
    \node[] (a33) at (4, 0)    {$a_{33}$};           
    \node[] (a13) at (0, 2)    {$a_{13}$};           

    \draw[norm] (a33) -- (a41);
    \draw[norm] (a41) -- (a62) ;
    \draw[norm] (a62) -- (a13) ;
    \draw[norm] (a24) -- (a51) ;
    \draw[norm] (a51) -- (a62) ;

  \end{tikzpicture}  
    \caption{The $E_6$ cluster}
    \label{fig:e6clust}
  \end{figure}
The knowledge of the symbol alphabet has been instrumental in the
computation of scattering amplitudes of six and seven particles by
making it possible to construct a finite ansatz made of all possible
words in the symbol \cite{Dixon:2011pw,Dixon:2011nj,Dixon:2013eka,Dixon:2014voa,Dixon:2014iba,Drummond:2014ffa,Dixon:2015iva,Caron-Huot:2016owq,Dixon:2016nkn}.

In \cite{Drummond:2017ssj,Drummond:2018dfd} it was observed that the possible iterated
discontinuities of hexagon and heptagon amplitudes are governed by a geometric
principle of cluster algebraic origin. Namely, for an amplitude to have a given pair of consecutive branch cuts, the corresponding symbol letters should appear together in at least one cluster in the cluster algebra ${\rm Conf}_n(\mathbb{P}^3)$. This property was referred to as \emph{cluster adjacency}. In more geometric terms the principle states that the boundary components of the cluster polytope that correspond to consecutive branch points labelled by the $a_{ij}$ are required to have a non-empty intersection. The adjacency relations imply the Steinmann relations \cite{Steinmann,Steinmann2,Caron-Huot:2016owq} on consecutive discontinuities of amplitudes. Interestingly, when combined with the initial entry condition imposing only physical branch cuts on the Euclidean sheet, the Steinmann relations (if imposed on every Riemann sheet \cite{Yorgosslides,Caron-Huot:2018dsv,DP}, also known as the extended Steinmann relations) conversely imply the adjacency conditions, at least up to the weights so far investigated.

Using the $E_6$ cluster in Figure \ref{fig:e6clust}, it is easy to
work out the letters that are allowed next to any $a_{ij}$. One can
freeze the node one is interested in and the remaining nodes together
with all their mutations populate its ``(homogeneous) neighbour set''
which we denote as $\hns{a_{ij}}$.
\begin{equation}
  \hns{a_{ij}} : \text{homogeneous combinations of letters that are cluster adjacent to $a_{ij}$}\, .
\end{equation}
Other neighbour sets can be obtained through cyclic rotations. For the
heptagon case the neighbour set relations are summarised in Table
\ref{fig:adjtable}.

We distinguish the two types of neighbour pairs. For some neighbour
pairs, there exists a cluster where they are connected by an arrow,
whereas some are nowhere connected by an arrow despite being allowed
neighbours. We call these pairs \emph{connected} and \emph{disconnected neighbours},
respectively.

Similarly, some pairs of letters, namely \emph{mutation pairs}, can be
obtained as mutations of each other and never appear in the same
cluster. While this is a sufficient condition for two letters to be
disallowed as neighbours, there are letters that never mutate to each
other but also never appear together in a cluster.

\begin{table}
  \centering
  {
  \setlength\tabcolsep{1.6pt}
\scalebox{1}{\begin{tabular}{r|ccccccc|ccccccc|ccccccc|ccccccc|ccccccc|ccccccc}
& \multicolumn{7}{|c}{\as{1}{i}} & \multicolumn{7}{|c}{\as{2}{i}} & \multicolumn{7}{|c}{\as{3}{i}} & \multicolumn{7}{|c}{\as{4}{i}} & \multicolumn{7}{|c}{\as{5}{i}} & \multicolumn{7}{|c}{\as{6}{i}} \\
\hline
\as{1}{1}& $\bcircle$& $\wcircle$& $\wcircle$& $\bdiamond$& $\bdiamond$& $\wcircle$& $\wcircle$& $\bdiamond$& $\bdiamond$& $\wcircle$& $\bcircle$& $\bdiamond$& $\bcircle$& $\wcircle$& $\bdiamond$& $\wcircle$& $\bcircle$& $\bdiamond$& $\bcircle$& $\wcircle$& $\bdiamond$& $\bcircle$& $\wcircle$& $\bdiamond$& $\wcircle$& $\wcircle$& $\bdiamond$& $\wcircle$& $\bcircle$& $\wcircle$& $\bdiamond$& $\wcircle$& $\wcircle$& $\bdiamond$& $\wcircle$& $\wdiamond$& $\bdiamond$& $\wcircle$& $\wcircle$&  $\wcircle$& $\wcircle$& $\bdiamond$\\
\as{2}{1}& $\bdiamond$& $\wcircle$& $\bcircle$& $\bdiamond$& $\bcircle$& $\wcircle$& $\bdiamond$& $\bcircle$& $\wcircle$& $\bcircle$& $\bdiamond$& $\bdiamond$& $\bcircle$&   $\wcircle$& $\bdiamond$& $\wcircle$& $\bdiamond$& $\bdiamond$& $\wcircle$& $\bdiamond$& $\bcircle$& $\bdiamond$& $\wcircle$& $\bdiamond$& $\wcircle$& $\bcircle$& $\bdiamond$& $\wcircle$& $\wcircle$& $\bdiamond$& $\bcircle$& $\wcircle$& $\bdiamond$& $\wcircle$& $\bdiamond$& $\wcircle$& $\bdiamond$& $\wcircle$& $\bcircle$& 
  $\wcircle$& $\bdiamond$& $\wcircle$\\
\as{3}{1}& $\bdiamond$& $\bdiamond$& $\wcircle$& $\bcircle$& $\bdiamond$& $\bcircle$& $\wcircle$& $\bdiamond$& $\bcircle$& $\bdiamond$& $\wcircle$& $\bdiamond$& $\bdiamond$&  $\wcircle$& $\bcircle$& $\wcircle$& $\bcircle$& $\bdiamond$& $\bdiamond$& $\bcircle$& $\wcircle$& $\bdiamond$& $\wcircle$& $\bdiamond$& $\bcircle$& $\wcircle$& $\bdiamond$& $\wcircle$& $\wcircle$& $\bdiamond$& $\wcircle$& $\bdiamond$& $\wcircle$& $\bcircle$& $\bdiamond$& $\wcircle$& $\wcircle$& $\bdiamond$& $\wcircle$& 
  $\bcircle$& $\wcircle$& $\bdiamond$\\
\as{4}{1}&$\bcircle$& $\wcircle$& $\bdiamond$& $\wcircle$& $\wcircle$& $\bdiamond$& $\wcircle$& $\bdiamond$& $\wcircle$& $\bdiamond$& $\bcircle$& $\wcircle$& $\bdiamond$& $\wcircle$& $\bdiamond$& $\wcircle$& $\bdiamond$& $\wcircle$& $\bcircle$& $\bdiamond$& $\wcircle$& $\bcircle$& $\wdiamond$& $\bdiamond$& $\wcircle$& $\wcircle$& $\bdiamond$& $\wdiamond$& $\bcircle$& $\wcircle$& $\wcircle$& $\wcircle$& $\wcircle$& $\wcircle$& $\wcircle$& $\wdiamond$& $\bdiamond$& $\wdiamond$& $\wcircle$& 
  $\wcircle$& $\wdiamond$& $\bdiamond$\\
\as{5}{1}& $\bcircle$& $\wcircle$& $\bdiamond$& $\wcircle$& $\wcircle$& $\bdiamond$& $\wcircle$& $\wcircle$& $\bdiamond$& $\wcircle$& $\bdiamond$& $\wcircle$& $\bcircle$& $\bdiamond$& $\wcircle$& $\bdiamond$& $\bcircle$& $\wcircle$& $\bdiamond$& $\wcircle$& $\bdiamond$& $\bcircle$& $\wcircle$& $\wcircle$& $\wcircle$& $\wcircle$& $\wcircle$& $\wcircle$& $\bcircle$& $\wdiamond$& $\bdiamond$& $\wcircle$& $\wcircle$& $\bdiamond$& $\wdiamond$& $\wdiamond$& $\bdiamond$& $\wdiamond$& $\wcircle$& $\wcircle$& $\wdiamond$& $\bdiamond$\\
\as{6}{1}& $\wdiamond$& $\bdiamond$& $\wcircle$& $\wcircle$& $\wcircle$& $\wcircle$& $\bdiamond$& $\wcircle$& $\wcircle$& $\bdiamond$& $\wcircle$& $\bcircle$& $\wcircle$& $\bdiamond$& $\wcircle$& $\bdiamond$& $\wcircle$& $\bcircle$& $\wcircle$& $\bdiamond$& $\wcircle$& $\wdiamond$& $\bdiamond$& $\wdiamond$& $\wcircle$& $\wcircle$& $\wdiamond$& $\bdiamond$& $\wdiamond$& $\bdiamond$& $\wdiamond$& $\wcircle$& $\wcircle$& $\wdiamond$& $\bdiamond$& $\bcircle$& $\wdiamond$& $\wcircle$& $\wdiamond$& $\wdiamond$& $\wcircle$& $\wdiamond$\\
\end{tabular}
}
}
  \caption{The adjacency relations of the ${\cal A}$ coordinates of ${\rm Conf}_7(\mathbb{P}^3)$. The symbols correspond to the following possible relations between different letters, as explained in the main text. $\bcircle$\,: connected neighbours, $\bdiamond$\,: disconnected neighbours, $\wcircle$\,: mutation pairs, $\wdiamond$\,: non-neighbours that never mutate into each other.}
  \label{fig:adjtable}
\end{table}

The concept of neighbour sets makes it natural to define the neighbour
set functions, namely the functions that end with letters that are in
the neighbour set of a given letter $\phi$. These functions are
particularly relevant for expressing any cluster-adjacency symbol of
weight $w$ as a $w-1,1$ coproduct:
\begin{equation}
  \label{eq:cacoproduct}
  f^{(k)}
  =
  \sum_{\phi_\alpha \in {\cal A}} \sum_k
  f_{\hns{\phi_\alpha},k} \otimes \phi_\alpha \, ,
\end{equation}
where the functions $f^{(w-1)}_{\hns{\phi_\alpha},k}$ enumerated by
the index $k$, form a basis for weight-($w-1$) neighbour set
functions. Note that the range of the index $k$ varies depending on
$\phi_\alpha$.

In Table \ref{neisets}, we reproduce the dimensions of the spaces in
which various types of heptagon neighbour-set functions with physical
branch cuts live. The dimensions of these spaces depend on the
letter the neighbours of which are allowed in the final entry. The
neighbour-set functions will play a central role in parameterising the
four-loop NMHV amplitude.  { \renewcommand{\arraystretch}{1.2}
  \begin{table}
    \centering
    \begin{tabular}{lllllll}
      \toprule
      Weight&2&3&4&5&6&7\\
      \midrule
      ${\rm hns}[a_{1i}]$&10&29&83&229&612&1577\\
      ${\rm hns}[a_{2i}]$ &15&43&117&311&804&2025\\
      ${\rm hns}[a_{4i}]$ &6&14&34&87&224&570\\
      ${\rm hns}[a_{6i}]$&4&11&29&76&193&476\\
      Full $E_6$&28&97&308&911&2555&6826\\
      \bottomrule
    \end{tabular}
    \caption{Dimensions of the neighbour-set function spaces of the
      heptagon alphabet with initial entries $a_{1i}$ and the dimensions
      of the full cluster-adjacent heptagon functions}
    \label{neisets}
  \end{table}
}

\subsection{R-invariants}
Cluster adjacency manifests itself not only in the symbols of
polylogarithms but also in the poles of R-invariants introduced in
\cite{Drummond:2008vq}.  R-invariants $[ijklm]$ are the basic invariants
of the Yangian symmetry \cite{Drummond:2009fd} of scattering amplitudes which combines both the superconformal and dual superconformal symmetries. When written in
twistor variables \cite{Mason:2009qx}, they depend on five twistors and are defined as
\begin{equation}\label{eq:R-invariants}
  [ijklm]
  =
  \frac{
    \delta^{0|4}(\chi^I_i \ab{jklm} + \text{cyclic})
  }{
    \ab{ijkl}\ab{jklm}\ab{klmi}\ab{lmij}\ab{mijk}
  }\,.
\end{equation}
Here the $\chi_i$ are Grassmann variables which complete the $Z_i$ into supertwistors. They transform in the fundamental representation of the $su(4)$ R-symmetry and encode all the different possible choices of NMHV component amplitudes which may be extracted from the NMHV super amplitude. We refer the reader to \cite{Drummond:2008vq,Mason:2009qx} for details on the structure of the supermultiplets and supertwistor variables.

The R-invariants satisfy six-term identities of the form
\begin{equation}
  [ijklm]
  - [ijkln]
  + [ijkmn]
  - [ijlmn]
  + [iklmn]
  - [jklmn]
  =
  0 .
  \label{eq:sixterm}
\end{equation}
Taking into account these identities as well as the identities among
the identities, the number of independent R-invariants for an
$n$-particle process is ${n-1 \choose 4}$. For the hexagon
and the heptagon, this counting gives 5 and 15 R-invariants
respectively. The BCFW recursion relations result in the following expression for
NMHV tree superamplitudes in terms of the R-invariants,
\begin{equation}
  \label{eq:nmhvbcfw}
  {\cal E}^{(0)}_{n,\text{NMHV}}=
  \sum_{2 < i < j\leq n}\bigl[1\,i-1\,i\,j-1\,j\bigr]\, .
\end{equation}

It was noted in \cite{Drummond:2018dfd} that each R-invariant
contains only cluster adjacent poles. This observation extends to the individual terms in
BCFW representations of tree amplitudes with higher degree of helicity
violation even though the terms therein contain more complicated
Yangian invariants as well as products of R-invariants.

\subsection{NMHV loop amplitudes and $\overline{Q}$ final entries}
\label{NMHVandQbar}

The observations above extend from individual symbols and R-invariants
to NMHV amplitudes in a way in which poles and symbol final entries
are related by cluster adjacency. We use this to
construct a simplified ansatz for the NMHV heptagon and comment in the
next section on how one can fix all its parameters using simple
physical constraints at four loops.

In the case of seven-particle scattering, there are 21 R-invariants, namely
\begin{equation}
  (12) = [34567]\,,
  \qquad
  (13) = [24567]\,,
  \qquad
  (14) = [23567]
\end{equation}
and their cyclic copies. They satisfy six independent six-term
identities making the number of independent R-invariants 15, which can be chosen as \cite{CaronHuot:2011kk}
\begin{equation}
  \label{eq:rinvbasis}
  \begin{aligned}
    &{\cal E}^{(0)}_{7,\text{NMHV}} = (12) + (14) + (34) + (16) + (36) + (56)\,,\\
    &(12) \qquad  \text{\& cyclic}\,,\\
    &(14) \qquad  \text{\& cyclic}\,.
  \end{aligned}
\end{equation}
This means that the $L$-loop NMHV heptagon amplitude can be written in the following form
\begin{equation}
  \label{eq:heptNMHVform}
  \mathcal{E}_{7,\rm NMHV}^{(L)} =
  E_0^{(L)}  {\cal E}^{(0)}_{7,\text{NMHV}} + \bigl(  E_{12}^{(L)} \, (12) + E_{14}^{(L)} \, (14) + \text{cyclic} \bigr),
\end{equation}
where $E_0, E_{12}$ and $E_{14}$ are all cluster adjacent
polylogarithms built on the heptagon alphabet (\ref{heptletters}). In this paper we use 
\begin{equation}
g^2=\frac{a}{2}=\frac{\lambda}{16\pi^2}
\end{equation}
as a loop-counting parameter, where $\lambda$ is the usual 't Hooft coupling.

To fully exploit the cluster adjacency in the final entries, we are
required to write an ansatz of a different form from (\ref{eq:heptNMHVform}), employing all 21
invariants. The $\overline{Q}$ equation \cite{CaronHuot:2011kk} already constrains the combinations of R-invariants times final entries that can appear in the amplitude, and by making use of the six-term
identities these can be recast in a manifestly cluster adjacent form \cite{Drummond:2018dfd}:
\begin{align}
  \hnsqbar{(12)} &= \{ a_{15}, a_{21}, a_{26}, a_{32}, a_{34}, a_{53}, a_{57}\}
                  \subset \hns{(12)}\notag \\
  \hnsqbar{(13)} &= \{ a_{21}, a_{23}, a_{31}, a_{33}, a_{41}, a_{43}, a_{62}\}
                  \subset \hns{(13)\notag }\\
  \hnsqbar{(14)} &= \{ a_{11}, a_{14}, a_{21}, a_{24}, a_{31}, a_{34}, a_{46}\}
                   \subset \hns{(14)}\qquad  \text{\& cyclic}\,,
                   \label{eq:qbarca}
\end{align}
where we introduced the notation $\hnsqbar{(ij)}$ to refer to such sets
henceforth. The above sets result in $3 \times 7 \times 7 = 147$ (final entry)$\otimes$(R-invariant) pairs which are allowed.

Note that, while these letters are cluster neighbours of all of the
poles of the corresponding R-invariant, the $\overline{Q}$ equation puts a
more stringent constraint on these (final entry)$\otimes$(R-invariant)
pairs. The poles of the R-invariants $(12)$ and $(13)$ would also
be compatible with the letters $\{a_{11},a_{12},a_{22},a_{31},a_{55}\}$ and $\{a_{12},a_{13}\}$,
respectively leading to a total of $(12+9+7)\times7 = 196$ cluster adjacent (final entry)$\otimes$(R-invariant) pairs. Of these the following 21,
\be
[(12)-(13)+(14)-(15)+(16)-(17)]\otimes\{a_{11},a_{21},a_{31}\}\quad \& \text{ cyclic,}
\ee
are identically zero due to identities. This leaves 175 independent cluster adjacent pairs, which the $\overline{Q}$ constraint reduces to the 147 shown in (\ref{eq:qbarca}).

As per the discussion above, we start with the following manifestly cluster adjacent and $\overline{Q}$ satisfying ansatz for
the $L$-loop BDS-like normalised NMHV amplitude,
\begin{equation}
  \label{eq:ansatz}
  \mathcal{E}_{7,\rm NMHV}^{(L)} =
  e_{12}^{(L)} \, (12) + e_{13}^{(L)} \, (13) + e_{14}^{(L)} \, (14) + \text{cyclic}.
\end{equation}
The $e_{ij}$ are tensor products of the form
\begin{equation}
  e_{ij}^{(L)} =
  \sum_{\phi_\alpha\, \in\, \hnsqbar{(ij)}}\,\,
   \sum_{k}
  \,\,
  c^{(ij)}_{k, \alpha}\,\, f^{(2L-1)}_{\hns{\phi_\alpha},k}  \otimes \phi_{\alpha}\,,
\end{equation}
conforming to the coproduct structure of cluster-adjacent functions described in
equation (\ref{eq:cacoproduct}) and with final entries $\phi_\alpha$
are chosen from the set $\hnsqbar{(ij)}$ defined in equation
(\ref{eq:qbarca}).

Note that adjacency (and $\overline{Q}$) helps in two ways in the
above ansatz. It reduces the possible final entries next to each
R-invariant and it also reduces the possible next-to-final entries
for a given final entry. This means that we do not even need a full
weight seven basis of cluster adjacent functions, we only need the
much smaller spaces whose final entries are compatible with each
$\phi_\alpha$ in turn.

We stress that the form (\ref{eq:ansatz}) is not unique due to the
six-term identities that the heptagon R-invariants satisfy and the
amplitude $\mathcal{E}^{(L)}$ needs to be integrable only on the support of
these identities. In order to obtain a manifestly integrable amplitude
one should express the 21 $(ij)$ in terms of a non-redundant set of 15,
e.g. those in equation (\ref{eq:rinvbasis}). In that basis, the
integrable coefficient functions are expressed in terms of
$e^{(L)}_{ij}$ as follows:
\begin{equation}\label{eq:Etoe}
  E^{(L)}_0   = \sum_{i=1}^{7} e^{(L)}_{i\,i+2},
  \qquad
  E^{(L)}_{14} = e^{(L)}_{14} - e^{(L)}_{16} - e^{(L)}_{46},
  \qquad
  E^{(L)}_{12} = e^{(L)}_{12}- e^{(L)}_{16} - e^{(L)}_{24} - e^{(L)}_{46}\, .
\end{equation}

It is possible to remove some redundancies of this ansatz using the
appropriate reflection symmetries of the coproducts
$e_{ij}^{(L)}$. For example $e_{12}^{(L)}$ is invariant under
$Z_i \mapsto Z_{3-i}$, which relates the terms ending with $a_{21}$,
$a_{26}$ and $a_{53}$ to those ending with $a_{32}$, $a_{34}$ and
$a_{57}$, respectively. Moreover, in $e_{12}^{(L)}$, $a_{15}$ is preceded
by a function which is invariant under the reflections of the twistors
that leave $Z_5$ invariant.

In the following section we will focus on the technical details
of the four-loop computation.

\section{The four-loop computation}
\label{sec-Comp}

We will first give an account of the free parameters
in the cluster-adjacent ansatz with dihedral symmetry
(\ref{eq:ansatz}) at four loops. We then describe the steps we took to
find the values of these parameters to determine the NMHV
amplitude. We also explain how one can use the ancillary files to
construct the symbol of the amplitude in explicit form.

Following the dimensions listed in Table \ref{neisets}, we can work
out the dimensions of \mbox{weight-$\{7,1\}$} tensor-product spaces in
which we are looking for the symbols $e_{ij}$. For example, consider the
neighbour-set functions associated with the seven final entries \(\{ a_{15}, a_{21}, a_{26}, a_{32}, a_{34}, a_{53}, a_{57}\}\) of the
symbol $e_{12}$, as
given in equation (\ref{eq:qbarca}). The weight-7 neighbour set
functions that come before $a_{15}$ live in a 1577-dimensional space,
those that come before $a_{21}$,$a_{26}$ $a_{32}$ and $a_{34}$ live in
a 2025-dimensional space and those that come before $a_{53}$ and
$a_{57}$ live in a 570-dimensional one. This amounts to a total of 10,817
unknown coefficients in the coproduct $e_{12}^{(L)}$ but taking the
reflection symmetry into account reduces this number to 5426. With a
similar counting, one has 4867 and 5919 unfixed coefficients for
$e_{13}^{(4)}$ and $e_{14}^{(4)}$, respectively, so that the
undetermined coefficients in (\ref{eq:ansatz}) number
16,212 in total. Requiring that $\cE_{7,{\rm NMHV}}^{(4)}$ is integrable, free of spurious
poles and has the correct collinear limits uniquely fixes all of these
coefficients.

We have implemented these constraints in separate stages. One can
start by requiring the integrability of the symbol
$E_{14} = e_{14} - e_{16} - e_{46}$. This leaves 8,444 unfixed
coefficients. Then one can impose the integrability of
$E_{12} = e_{12}- e_{16} - e_{24} - e_{46}$ bringing this number down
to 56. Once the integrability of $E_{12}$ and $E_{14}$ is imposed,
there are no new constraints coming from the integrability of
$E_0$. In this 56-dimensional space, one can then look for
combinations for which the amplitude is free of spurious poles. These
are poles that could potentially appear in the limit where one of the
4-brackets in the denominator of the R-invariants vanishes. However in
physical amplitudes such poles are only allowed when the 4-bracket is
of the form $\langle i-1 i j-1 j\rangle$, corresponding to an
intermediate particle becoming on shell. In all other cases, this
potential pole must be cancelled by a vanishing of its transcendental
component, which, after also taking into account cyclic symmetry,
implies the following conditions:
\begin{align}\label{eq:spurious_poles}
\text{Spurious I:} \quad  E_{47}|_{\langle 1356 \rangle =0} &= 0\,, \\
\text{Spurious II:} \quad  E_{23}|_{\langle 1467 \rangle =0} &= E_{25}|_{\langle 1467 \rangle =0}\,,
\end{align}
which have been worked out in
\cite{Korchemsky:2009hm,Dixon:2016nkn}. Imposing both conditions
described in equation (\ref{eq:spurious_poles}), one is left with only
five coefficients to be determined by imposing a kinematic limit, such
as the collinear limits.

In the collinear limit, two of the neighbouring particles in a
colour-ordered amplitude become proportional to each other with an
unspecified proportionality constant. We follow \cite{Dixon:2016nkn}
to describe the collinear limit in the momentum twistor.  In a generic
configuration, the momentum twistor $Z_7$ can be parametrised as a
linear combination of four other momentum twistors as follows:
\begin{equation}
  Z_7
  =
  Z_1
  +
  \epsilon
  \frac{
    \ab{1456}
  }{
    \ab{2346}
  }
  Z_2
  +
  \tau
  \frac{
    \ab{1245}
  }{
    \ab{2456}
  }
  Z_6
  +\eta
  \frac{
    \ab{1256}
  }{
    \ab{2456}
  }
  Z_4\, .
\end{equation}
A collinear configuration is obtained when one sends first
$\eta \to 0$ followed by $\epsilon \to 0$. The parameter $\tau$ then
relates the momentum fraction.

Scattering amplitudes in planar ${\cal N}=4$ super Yang-Mills have a
well-known collinear behaviour and they can be related to the
amplitude with one fewer particle. Usually the BDS-normalised amplitude
\cite{Anastasiou:2003kj,Bern:2005iz}
 ${\cal B}_n$ is used to consider collinear kinematics, as opposed to
the BDS-like normalised one, because the former is finite in this
limit and directly reduces to the quantity of one fewer particle
${\cal B}_{n-1}$. The two quantities are related via
\be
\mathcal{B}_n = \exp{\biggl(-\frac{\Gamma_{\text{cusp}}}{4}\, Y_n \biggr)} \mathcal{E}_n,\qquad Y_n \equiv - \mathcal{E}^{(1)}_{n,{\rm MHV}}\,.
\label{eq:E7componentBDSToBDSlike}
\ee
In other words a BDS-like normalised
quantity ``$\cE$'', which may be the full superamplitude \eqref{eq:BDS-like_normalisation}, or a given MHV sector $\mathcal{E}_{{\rm N}^k{\rm MHV}}$ thereof \eqref{eq:BDS-like_superamplitude}, or a particular transcendental component of the latter such as $E_0, E_{ij}$ in \eqref{eq:heptNMHVform}, is related to the corresponding BDS-normalised quantity ``$\cB$''  by an exponential factor involving the one-loop MHV amplitude and the cusp anomalous dimension $\Gamma_{\rm cusp}$.

Explicitly for $n=6,7$, the functions $Y_n$ are given by:
\begin{align}
Y_6 &= -\sum_{i=1}^3 \biggl[ \text{Li}_2\left(1-\frac{1}{u_i}\right)\biggr]\,, \notag \\
Y_7 &= -\sum_{i=1}^7 \biggl[ \text{Li}_2\left(1-\frac{1}{u_i}\right) +
\frac{1}{2} \log
\left(\frac{u_{i+2}u_{i{-}2}}{u_{i+3}u_{i}u_{i{-}3}}\right) \log u_i
\biggr]\,,
\label{Ydef}
\end{align}
where $u_i = u_{i,i+3}$ in terms of the cross-ratios defined in eq. (\ref{eq:xratios}).

For seven particle scattering, there are two types of combinations of
the NMHV superamplitude components that produce six-point
amplitudes. The ``$k$-decreasing'' combination of BDS-normalised
functions produces the six-point MHV superamplitude, whereas the
``$k$-preserving'' ones produce the five independent components of the six-point NMHV superamplitude \cite{Drummond:2014ffa}.

For example, in the $k$-decreasing collinear limit, the MHV ($\xi^0$)
component of the six-particle amplitude receives contribution from a
number of functions multiplying different R-invariants. More
precisely, the combination 
\begin{equation}
  B_0 + B_{23} + B_{34} = \exp{\biggl(-\frac{\Gamma_{\text{cusp}}}{4}\, Y_7 \biggr)}\,
  \bigl(E_0 + E_{23} + E_{34}\bigr)\, ,
\end{equation}
where $\Gamma_{\text{cusp}}$ is the cusp anomalous dimension, is what
is expected to reproduce the BDS-normalised six-point MHV
amplitude $\mathcal{B}_{6,{\rm MHV}}$.

Especially at four loops, it is cumbersome to compute the
BDS-normalised functions, which contain redundant information in that
they involve a large number of known products. However, with the
knowledge of the MHV heptagon amplitude, it is not necessary to
convert between the different normalisations of the
amplitude. Instead, one can consider the difference
\begin{equation}
  \label{eq:colldiff1}
  E_0 + E_{23}+E_{34} - {\cal E}_{7,\rm MHV} \, ,
\end{equation}
where ${\cal E}_{7,\rm MHV}$ is the known BDS-like normalised
heptagon amplitude. Since $ {\cal E}_{7,\rm MHV}$ and the
combination \(E_0 + E_{23}+E_{34} \) both reduce to the same quantity
\begin{equation}
  E_0 + E_{23}+E_{34}\,,\,\,
  {\cal E}_{7,\rm MHV}
  \,\,
  \rightarrow
  \,\,
  \biggl[
  \exp{\biggl(\frac{\Gamma_{\text{cusp}}}{4} (Y_6-Y_7)}\biggr)\,
  {\mathcal{E}_{6,{\rm MHV}}}
  \biggr]\biggr|_L
  \qquad 
  \text{(collinear limit)}
\end{equation}
in the collinear limit, where $\mathcal{E}_{6,\rm MHV}$ is the six
particle amplitude, one can impose the vanishing of the difference
(\ref{eq:colldiff1}) which only contains relatively simple,
cluster-adjacenct quantities.

While the vanishing of (\ref{eq:colldiff1}) in the collinear limit is
a sufficient constraint to uniquely fix the amplitude, constraints
that relate ${\cal E}^{(4)}_{7,{\rm NMHV}}$ to another amplitude are
not strictly necessary. One can still explicitly construct the
BDS-normalised amplitude in either $k$-decreasing or $k$-preserving
collinear limit and determine ${\cal E}^{(4)}_{7,{\rm NMHV}}$ only by
requiring the finiteness of the limit, without prior knowledge of
${\cal E}^{(4)}_7$.

Integrability and the cancellation of spurious poles are linear
constraints on the space of coproducts and finding their solution
spaces can be formulated as null-space problems for integer-valued
matrices encoding these constraints (see \cite{Dixon:2016nkn} for
details). We found that an efficient way of computing these kernels is
to work modulo a prime number and feed the constraint matrices into
SpaSM \cite{spasm}, a sparse linear solver that employs modular arithmetic. One can
then compute these null-spaces modulo $p$ for various prime numbers
$p$ and reconstruct the exact amplitude using the Chinese Remainder
Theorem. However, it was possible to guess the answer that satisfies
all constraints exactly by only repeating the calculation mod 43051
and mod 46153.

\subsection{Explicit results}
The explicit tensor products $e^{(4)}_{ij}$ are too large to be
included as supplementary material to this paper and therefore we
provide them encoded as \{6,1,1\} coproducts. In this section we
describe how one can use the provided data to reconstruct the
amplitude.

The file \verb+e74.m+ contains a 4-index tensor
of dimensions $3\times 2555 \times 42 \times 42$ in \verb+Mathematica+
\verb+SparseArray+ format. The first index enumerates the tensors
$e^{(4)}_{12}$, $e^{(4)}_{13}$ and $e^{(4)}_{14}$. Once the first
index is specified, the remaining array contains the coefficients
$c_{ij}^{k\alpha\beta}$ in the coproduct representation of
$e^{(4)}_{ij}$ :
\begin{equation}
  \label{eq:eijcoproduct}
  e^{(4)}_{ij}
  =
  \sum_{k=1}^{2555}\,\sum_{\alpha=1}^{42}\,\sum_{\beta=1}^{42}
  \,\,c_{ij}^{k\alpha\beta}\,\,
  f_k^{(6)} \otimes \phi_\alpha \otimes \phi_\beta\, .
\end{equation}

We also provide bases spanning the spaces of weight-$w$ $f_k^{(w)}$
in terms of \{$w-1$,1\} coproducts:
\begin{equation}
  f_k^{(w)}
  =
  \sum_{\ell}^{\dim_{w-1}}\,\sum_{\alpha}^{42}
  \,\, M^{(w)}_{k \ell \alpha}\,\,
  f_{\ell}^{(w-1)} \otimes \phi_\alpha\, ,
\end{equation}
using which one can recursively construct the symbols of functions
$f^{(w)}_k$ in order to convert the coproducts (\ref{eq:eijcoproduct})
to symbols. The coefficients $M^{(w)}_{k \ell \alpha}$ are encoded in
the files \verb+m+$w$\verb+.m+ as \verb+SparseArray+ objects for $2 \leq w \leq 6$.

Due to the cyclic symmetry of the superamplitude, the coefficients
$c_{1j}^{k\alpha\beta}$ for $j =2,3,4$ are sufficient to describe the
amplitude. The remaining symbols $c_{ij}^{k\alpha\beta}$ with
$i \neq 1$ can be constructed by rotating the coproduct form
(\ref{eq:eijcoproduct}). For example $e^{(4)}_{23}$ can be constructed
as
\begin{equation}
  e^{(4)}_{23}
  =
  \sum_{k=1}^{2555}\,\sum_{\alpha=1}^{42}\,\sum_{\beta=1}^{42}
  \,\,c_{12}^{k\alpha\beta}\,\,
  {\cal C} \bigl[ f_k^{(6)}\bigr]
  \otimes
  {\cal C} \bigl[\phi_\alpha \bigr]
  \otimes
  {\cal C} \bigl[\phi_\beta \bigr]\, ,
\end{equation}
where the cyclic rotation operator ${\cal C}$ acts on the letters as
\begin{equation}
    {\cal C} \bigl[a_{i\,j} \bigr] = a_{i\,j+1} 
\end{equation}
while its action on the functions is a linear transformation
in the corresponding function space:
\begin{equation}
  {\cal C} \bigl[ f_k^{(w)}\bigr]
  =
  \sum_{\ell=1}^{\text{dim}_w}
  {\cal C}^{(w)}_{k\ell} \, f_\ell^{(w)}\, .
\end{equation}
The matrices ${\cal C}^{(w)}_{k\ell}$ are given as a Mathematica
\verb+List+  in the file \verb+rotationmatrices.m+ for $2\leq w \leq 6$.

Following this procedure one obtains three 8-dimensional Mathematica
\verb+SparseArray+ objects encoding the symbols of $e^{(4)}_{ij}$
which enter the coefficient functions $E_*^{(4)}$. These then can be
used to perform various analyses of our result, such as the
investigation of its behaviour in the multi-Regge kinematics.

\section{Multi-Regge limit}
\label{sec-MRK}

In this section, we will consider the multi-Regge limit of our $n=7$, 4-loop NMHV symbol, with a two-fold aim: First, to check our calculation against independent results available for the amplitude in this limit up to NLLA \cite{DelDuca:2016lad,DelDuca:2018hrv}. And second, to obtain new predictions up to N$^3$LLA, which we hope will play an important role in further elucidating the perturbative structure of the heptagon in the limit, and guide its finite-coupling determination, similarly to the hexagon case. We start by briefly reviewing the kinematics and the most natural amplitude normalisation for our purposes in subsections \ref{sec:MRKinematics} and \ref{sec:BDS+contunuation}, before describing the evaluation of the amplitude in subsection \ref{sec:MREvaluation}. The reader interested in the final result and comparison may just skip directly to subsection \ref{sec:MRPredictions}.

\subsection{Kinematics}\label{sec:MRKinematics}
We will focus on $2\to 5$ scattering, for which multi-Regge kinematics (MRK) is defined as the limit where the produced particles are strongly ordered in rapidity. For $\cN=4$ SYM, the nontrivial kinematic dependence is encoded in dual conformal cross ratios, and in \cite{Bartels:2012gq,Bartels:2014mka} it was shown that in the following convenient choice of six algebraically independent cross ratios, the limit becomes\footnote{In the notations of \cite{DelDuca:2016lad,DelDuca:2018hrv}, $u_{i+1,j_1}=U_{ij}$, due to different numbering of momentum twistors.}
\be\label{eq:UcrossratiosMRK}
v_{1i}\equiv u_{i+2i+5}\to 1\,,\quad  v_{2i}\equiv u_{1i+3}\to 0\,,\quad v_{3i}\equiv u_{2i+4}\to 0\,,
\ee
with
\be
\frac{v_{2i}}{1-v_{1i}}\equiv \frac{1}{|1-z_i|^2}\,,\quad \frac{v_{3i}}{1-v_{1i}}\equiv \frac{|z_i|^2}{|1-z_i|^2}\,,\quad i=1,2\,,
\ee
held fixed. The right hand-side defines the four real, or two complex, finite cross ratios $z_1$, $z_2$ that parametrise the limit, whereas the two cross ratios that become infinitesimal may be chosen as
\be\label{eq:tau_def}
\tau_i\equiv\sqrt{v_{2i}v_{3i}}\,,\quad i=1,2\,.
\ee

From the above equations, we may also deduce the behaviour of all heptagon symbol letters, eq.\eqref{heptletters}, in MRK, which is a necessary step before evaluating the corresponding amplitude. Let us therefore record it here before closing this subsection,
{
\be\label{eq:aMRK}\footnotesize
\begin{aligned}
a_{14}&\to \frac{1}{a_{12}} & a_{15}&\to a_{11} a_{12} & a_{16}&\to \frac{1}{a_{11}} & a_{24}&\to \frac{a_{13}}{a_{23}}  & a_{27}&\to \frac{a_{17}}{a_{21}}\\
 a_{32}&\to \frac{a_{12} a_{13}}{a_{23}} & a_{33}&\to \frac{a_{23}}{a_{12}} & a_{36}&\to \frac{a_{21}}{a_{11}} & a_{37}&\to \frac{a_{11} a_{17}}{a_{21}} & a_{41}&\to \frac{a_{23} a_{26}}{a_{12}} \\
 a_{42}&\to \frac{a_{17} a_{34}}{a_{21}} & a_{43}&\to a_{11} a_{23}  & a_{44}&\to \frac{a_{34}}{a_{11}} & a_{45}&\to \frac{a_{11} a_{17} a_{23}}{a_{21}} & a_{46}&\to \frac{a_{26}}{a_{12}}\\
 a_{47}&\to \frac{a_{11} a_{12} a_{17}}{a_{21}} & a_{51}&\to \frac{a_{13} a_{35}}{a_{23}} & a_{52}&\to \frac{a_{21} a_{25}}{a_{11}} & a_{53}&\to \frac{a_{11} a_{12} a_{13}}{a_{23}} & a_{54}&\to \frac{a_{25}}{a_{11}} \\
 a_{55}&\to \frac{a_{12} a_{13} a_{21}}{a_{23}} & a_{56}&\to \frac{a_{35}}{a_{12}} & a_{57}&\to a_{12} a_{21} & a_{61}&\to a_{12} a_{17} &a_{62}&\to a_{11} a_{13}  \\
 a_{63}&\to \frac{a_{25} a_{34}}{a_{11}} & a_{64}&\to a_{11} a_{12} a_{13} & a_{65}&\to \frac{1}{a_{11} a_{12}} & a_{66}&\to a_{11} a_{12} a_{17} & a_{67}&\to \frac{a_{26} a_{35}}{a_{12}}\,,&&
\end{aligned}
\ee
}
where we see that only 12 out of the 42 letters remain multiplicatively independent in the limit. These 12 letters may in turn be expressed in terms of the variables \eqref{eq:UcrossratiosMRK}-\eqref{eq:tau_def} as
{
\be\label{eq:aTozMRK}\footnotesize
\begin{aligned}
 a_{11}&\to \frac{\tau _1}{z_2  \bar{z}_2 \sqrt{z_1\bar{z}_1} } & a_{12}&\to \tau _2 z_1 \bar{z}_1 \sqrt{z_2 \bar{z}_2} & a_{13}&\to \frac{\sqrt{z_2 \bar{z}_2}}{\tau _1^2 \tau _2}\\
a_{17}&\to \frac{1}{\tau _1 \tau _2^2\sqrt{z_1 \bar{z}_1}} & a_{21}&\to -\frac{1}{\tau _2 z_1 \sqrt{z_2 \bar{z}_2}}& a_{22}&\to\frac{\tau _1 \tau _2 \left(\bar{z}_2-\bar{z}_1 \bar{z}_2-1\right)}{\sqrt{z_2 \bar{z}_2}}\sqrt{\frac{z_1}{\bar{z}_1}}\\
a_{23}&\to -\frac{z_2\sqrt{z_1 \bar{z}_1}}{\tau _1} & a_{25}&\to -\frac{\tau _1\left(1-z_1\right)}{\bar{z}_2\sqrt{z_1 \bar{z}_1}} & a_{26}&\to \tau _2\bar{z}_1 \left(1-z_2\right)  \sqrt{\frac{\bar{z}_2}{z_2}}\\
a_{31}&\to\frac{\tau _1 \tau _2\left(z_2-z_1 z_2-1\right) }{\sqrt{z_2 \bar{z}_2}} \sqrt{\frac{\bar{z}_1}{z_1}}& a_{34}&\to -\frac{\tau _1 \left(1-\bar{z}_1\right)}{z_2\sqrt{z_1 \bar{z}_1}} & a_{35}&\to \tau _2 z_1 \left(1-\bar{z}_2\right)\sqrt{\frac{z_2}{\bar{z}_2}} \,.\\
\end{aligned}
\ee
}

\subsection{BDS normalisation and analytic continuation}\label{sec:BDS+contunuation}

While in the previous sections we determined the heptagon superamplitude in the
BDS-like normalisation\footnote{In what follows, we will drop the particle multiplicity index $n$, since we will be focusing on $n=7$.} $\mathcal{E}_{\rm NMHV}$, in MRK it is most
conveniently described in the BDS normalisation, introduced in (\ref{eq:E7componentBDSToBDSlike}).
In this paper we are exclusively dealing with symbols, and since
$\cS (\Gamma_{\text{cusp}})=4g^2$, we may write each term in the weak
coupling expansion of \eqref{eq:E7componentBDSToBDSlike} at symbol
level as
\begin{equation}
  \label{eq:BDSLikeToBDS}
  \cE^{(L)}
  =
  \sum_{k=0}^L \cB^{(k)} \frac{(-Y)^{L-k}}{(L-k)!} \,.
  \end{equation}
Given the conformal equivalence of MRK with the double soft limit for the heptagon, all loop corrections to the corresponding BDS-normalised superamplitude will vanish in the Euclidean region. In order to obtain a nontrivial result, we therefore need to analytically continue the latter amplitude to different kinematic regions, and here we will chose the region where we analytically continue the energy components of all produced particles to opposite sign. In terms of the conformally invariant cross ratios, this amounts to
\be\label{eq:Ucontinuation}
u_{73}\overset{\cC}{\to} e^{-2\pi i}u_{73}\,,
\ee
and given the relation of the latter to the $a_{1i}$ letters,
\be
a_{11}= \frac{u_{14} u_{51}}{u_{36} u_{62} u_{73}}\quad+\,\,\text{cyclic}\,,
\ee
it is evident that the amplitude in this region will differ from its value in the Euclidean region by
\begin{align}\label{discB}
\Delta \cB\equiv \cB^{\mathcal{C}}-\cB=&-2\pi i\text{Disc}_{u_{73}}\cB\\
=&-2\pi i\left(-\text{Disc}_{a_{11}}\cB-\text{Disc}_{a_{12}}\cB+\text{Disc}_{a_{13}}\cB-\text{Disc}_{a_{15}}\cB+\text{Disc}_{a_{17}}\cB\right)\,.\nonumber
\end{align}
Note that the above equation also holds for each component of the superamplitude separately, since the R-invariants \eqref{eq:R-invariants} are rational functions of the kinematics, and thus they will remain unchanged under the analytic continuation.

\subsection{Evaluating the gluon amplitude in the limit}\label{sec:MREvaluation}
So far our discussion was at the level of the entire superamplitude, however in MRK the natural object to consider are its gluon amplitude components, since the theoretical framework for describing the limit was born out of the study of strong interactions. Focusing on $1+2\to3+\ldots+7$ scattering in all-outgoing momenta conventions, and denoting the helicities of the produced gluons as $h_1,h_{2},h_{3}$, without loss of generality can define the relevant BDS-normalised gluon amplitudes as
\be\label{eq:Rhhh_definition}
\cR_{h_1,h_{2},h_{3}} \equiv \frac{A(-,-,+,h_1,h_{2},h_{3},+)}{A^{\textrm{BDS}}(-,-,+,h_1,h_{2},h_{3},+)}\Big|_{\text{MRK}}\,,
\ee
since the high energy of the incoming gluons implies that helicity is preserved along their lines in the limit. Particularly for the NMHV case, which is the focus of this paper, there exist two inequivalent helicity configurations, $\cR_{-++}$ and $\cR_{+-+}$, since $\cR_{++-}$ may be obtained from the former by a discrete parity and target-projectile (a particular dihedral flip that commutes with the limit) transformation.

The gluon amplitudes \eqref{eq:Rhhh_definition} can be extracted from the superamplitude, as coefficients of particular monomials of the fermionic variables $\chi^I_i $ entering in the R-invariants \eqref{eq:R-invariants}, with the latter being polynomials in these variables due to the fermionic delta function. We defer the details of the calculation to a future publication \cite{HeptLimits}, and just quote the final answer\footnote{The result of this calculation was also reported in \cite{DelDuca:2016lad}, but with the $\hat B_{ij}$ components cyclically permuted up by two compared to here, as a result of inconsistent conventions for momentum twistors.}, 
\begin{align}
\cR_{-++}=\hat{B_0}&+\hat{B}_{67}+\hat{B}_{71}+R_{234}\Big(\hat{B}_{51}-\hat{B}_{71}\Big)+R_{235}\Big(\hat{B}_{56}-\hat{B}_{51}\Big)\,,\notag\\ 
\cR_{+-+}=\hat{B_0}&+\hat{B}_{14}+\hat{B}_{47}+\hat{B}_{73}+\overline{R}_{234}\Big(\hat{B}_{12}-\hat{B}_{14}-\hat{B}_{47}\Big)\notag\\
&+R_{345}\Big(\hat{B}_{36}-\hat{B}_{14}\Big)+ \overline{R}_{234}R_{345}\Big(\hat{B}_{14}+\hat{B}_{62}-\hat{B}_{12}\Big)\,,
\label{RmppRpmpwrtB}
\end{align}
expressing the gluon amplitude as a linear combination of the independent, integrable components of the BDS-normalised NMHV superamplitude $\hat B_\ast$ with index $\ast$ equal to 0 or $ij$, after we analytically continue it and take its multi-Regge limit, times the independent rational factors coming from the R-invariants in the limit,
\be\label{eq:LeadingSing}
R_{234}=-\frac{z_1}{1-z_1}\,,\quad R_{235}=\frac{z_1 z_2}{1-z_2+z_1 z_2}\,,\quad R_{345}=-\frac{z_2}{1-z_2}\,,
\ee
with the corresponding barred quantities being their complex conjugates.

In principle we now have everything laid out for extracting the symbol of the 4-loop NMHV gluon amplitudes in MRK from the corresponding superamplitude in general kinematics, however in the current order the computation requires the tedious step of converting from the BDS-like to BDS normalisation in general kinematics, eq.~\eqref{eq:BDSLikeToBDS}. Instead, we have found it significantly more efficient to obtain the final result directly from the discontinuity of $E$ as follows: From the definition in the first line of \eqref{discB}, it is evident that the discontinuity of a product of symbols $F,G$ obeys the Leibniz rule,
\be
\Delta (F\cdot G)=(F+\Delta F) \cdot (G+\Delta G)-(F\cdot G)=\Delta F\cdot G+ F \cdot \Delta G\,,
\ee
since the $\Delta F \cdot \Delta G$ term has an additional factor of $\pi$, and is thus beyond the symbol. With the help of this property, and eq.~\eqref{eq:BDSLikeToBDS}, is is straightforward to relate the discontinuities of the symbols of the BDS and BDS-like amplitudes in MRK,
\be\label{eq:DBDSLikeToDBDSMRK}
  \Delta \hat \cE^{(L)}=\sum_{k=1}^L \left(\hat \cB^{(k)}-\delta_{k1}\Delta \hat Y\right) \frac{(-\hat Y)^{L-k}}{(L-k)!} \,,
\ee
where also we also used the fact that in MRK $\cB^{(k)}
\to\delta_{k0}$ before analytic continuation, and thus $\Delta \hat \cB^{(k)}=\hat \cB^{(k)}$ for $k\ge 1$. In the above relation, the function $Y$ and its discontinuity evaluate in the limit to
\be
\begin{aligned}
\hat Y=&2 (\log ^2 \tau _1 + \log ^2 \tau _2 + \log  \tau _1  \log  \tau _2) + \log  \tau _2\log  |z_1|^2 -\log  \tau _1 \log  |z_2|^2\\
&+\frac{1}{2} (\log ^2 |z_1|^2+\log ^2 |z_2|^2+\log |z_1|^2\log |z_2|^2)\\
 \Delta \hat Y=&-2\pi i\left[ -2 \log  \tau _1 -2 \log  \tau _2 +\log  |z_2|^2 -\log  |1-z_2+z_1 z_2|^2\right]\,, 
\end{aligned}
\ee
and $|z|^2=z\bar z$ etc. 

This completes our method for obtaining the BDS-normalised gluon amplitudes \eqref{eq:Rhhh_definition}, focusing on the NMHV configurations $\cR_{-++}$ and $\cR_{+-+}$. To summarise, starting with the symbol of $\mathcal{E}_{\rm NMHV}^{(L)}$, eq.~\eqref{eq:ansatz},  we take the linear combinations of its transcendental components, eq.~\eqref{eq:Etoe}, that appear in the right-hand side of \eqref{RmppRpmpwrtB} upon replacing $\hat B_*\to E_*$. For each such component, we take its discontinuity as in the second line of \eqref{discB}, and then its multi-Regge limit as in \eqref{eq:aMRK} and \eqref{eq:aTozMRK}, sequentially. Finally, we plug the result on the left-hand side of \eqref{eq:DBDSLikeToDBDSMRK}, which is valid not only for the entire superamplitude, but also for its components separately, and solve for $\hat \cB^{(L)}$ recursively, starting from $L=1$. For example, at 4 loops we will have
\be
 \hat \cB^{(4)}= \Delta \hat \cE^{(4)} + \frac{1}{6} \hat Y^3 \left(\hat \cB^{(1)}- \Delta \hat Y\right)-\frac{\hat Y^2  }{2}\hat \cB^{(2)} +\hat Y  \hat \cB^{(3)}\,.
\ee

\subsection{Comparison with BFKL approach and new predictions}\label{sec:MRPredictions}

In this final subsection, we will compare our findings for the 4-loop NMHV heptagon in MRK with independent results obtained for the latter to LLA \cite{DelDuca:2016lad} and NLLA \cite{DelDuca:2018hrv}, based on the Balitsky-Fadin-Kuraev-Lipatov (BFKL) approach \cite{Bartels:2011ge,Bartels:2013jna,Bartels:2014jya}. We will also discuss our new predictions for the amplitude in question up to N$^3$LLA.

Let us start by reviewing what has been previously known for the heptagon in the limit. Building on earlier work at LLA, in \cite{DelDuca:2018hrv} an all-loop dispersion integral was presented, yielding the $2\to 5$ amplitude in MRK to arbitrary logarithmic accuracy. It reads,
\be
\label{eq:Rhhhstart}
 \cR_{h_1h_2h_{3}}e^{ i \delta_7(z_1,z_2)} =2\pi i f_{h_1h_2h_3}\,,
\ee
where the right-hand side has the form of a Fourier-Mellin transform,
\begin{align}
f_{h_1h_2h_3}= &\frac{a}{2} \sum_{n_1,n_2=-\infty}^\infty 
\left(\frac{z_1}{\bar z_1}\right)^{\frac{n_1}{2}} \left(\frac{z_2}{\bar z_2}\right)^{\frac{n_2}{2}}
\int \frac{d\nu_1 d\nu_2}{(2\pi)^2}
   |z_1|^{2i\nu_1}|z_2|^{2i\nu_2}\tilde \Phi(\nu_1,n_1)\tilde \Phi(\nu_2,n_2)
\nonumber\\
&\times e^{-L_1 \omega(\nu_1,n_1)-L_2\omega(\nu_2,n_2)}I^{h_1}(\nu_1,n_1)\tilde C^{h_2}(\nu_1,n_1,\nu_2,n_2)
 \bar I^{h_3}(\nu_2,n_2)\,, \label{eq:f_hhh}
\end{align}
with
\be
L_i=\log\tau_i+i\pi\,, \quad \delta_7(z_1,z_2) =\frac{\pi\Gamma_{\text{cusp}}}{4}\log\frac{|z_1z_2|^2}{|1-z_2+z_1z_2|^4}\,,
\ee
which we see evidently depends on the variables  \eqref{eq:UcrossratiosMRK}-\eqref{eq:tau_def} that naturally describe the limit. Following the conventions of the relevant literature, in this section we have also switched our coupling normalisation to
\be
a=2g^2\,.
\ee

The remaining quantities in the integral \eqref{eq:f_hhh} are associated to the effective particle whose exchange governs the multi-Regge limit, known as the reggeised gluon or reggeon. In the kinematic region characterised by the analytic continuation \eqref{eq:Ucontinuation}, we have in particular a two-reggeon bound state, whose energy is the BFKL eigenvalue $\omega(\nu,n)$, and whose creation (annihilation) with a simultaneous emission a new final-state gluon of helicity $h_1$ ($h_3$) is encoded in the combined quantity $\tilde \Phi I^{h_1}$ ($\tilde \Phi \bar I^{h_2}$) known as the BFKL impact factor. These building blocks also appear in the hexagon amplitude, and they can be determined from first principles at weak coupling~\cite{Bartels:2008sc, Fadin:2011we}, or even to all loops with the help of integrability \cite{Basso:2014pla}.

Finally, the genuinely heptagonal quantity $\tilde C^{h_2}$, known as the central emission vertex, describes the emission of a gluon of helicity $h_2$ from the reggeon bound state in the middle of the ladder\footnote{In \eqref{eq:f_hhh}, the impact factor and central emission block have been rescaled compared to their original definition, so as to better expose their analytic properties, but this does not alter their physical interpretation.}. It was originally determined at leading order in \cite{Bartels:2011ge}, whereas its next-to-leading order correction was extracted from the 2-loop MHV heptagon, after promoting its known symbol \cite{Bargheer:2015djt} to a function, in \cite{DelDuca:2018hrv}. Plugging this correction back to the integral \eqref{eq:f_hhh} it is then possible to compute the amplitude at higher loops to NLLA, and this was indeed carried out for $\cR^{(4)}_{-++}$. In more detail, at weak coupling the amplitude in MRK also has a natural expansion in large logarithms in the infinitesimal $\tau_i$ variables, whose perturbative coefficients may be defined as
\begin{align}\label{eq:perturbativeExpansion}
\cR_{h_1,h_2,h_3}\left(\tau_{1},{z_{1}},\tau_{2},{z_{2}}\right)e^{i\delta_7(z_1,z_2)} &=1\,+ 2\pi i \sum_{\ell=1}^{\infty}\sum_{i_1,i_{2}=0}^{\ell-1}a^\ell \,\left(\prod_{k=1}^{2}\frac{1}{i_k!}\log^{i_k}\tau_k\right) \\
&\times \left( \,{\tilde g}_{h_1,h_2,h_{3}}^{(\ell;i_1,i_2)}(z_1,z_{2}) + 2 \pi i \, {\tilde h}_{h_1,h_2,h_{3}}^{(\ell;i_1,i_{2})}(z_1,z_{2}) \right)\,. \nonumber
\end{align}
The maximal logarithmic order amounts to $i_1+i_2=\ell-1$, as a consequence of the fact that all building blocks of the integrand \eqref{eq:f_hhh} start at $\cO(1)$, except for $\omega(\nu,n)$, which starts at $\cO(a)$. These coefficients have already been determined in \cite{DelDuca:2016lad}, in the notation
\be
{\tilde g}_{h_1,h_2,h_{3}}^{(\ell;i_1,i_2)}\to {g}_{h_1,h_2,h_{3}}^{(i_1,i_2)}=\texttt{LL[\{i1,i2\},\{h1,h2,h3\}]}\,,
\ee
where we also provided their naming in the ancillary files \texttt{NMHVLL7.m} and \texttt{NMHVLL6.m} accompanying the paper in question. The latter file is needed because of the interesting factorisation property
\be\label{eq:factorization}
{g}_{-++}^{(i_1,0)}(\rho_1,\rho_{2})={g}_{-+}^{(i_1)}(\rho_1)\,,
\ee
reducing heptagon perturbative coefficients to hexagon ones, after one first expresses them in so-called simplicial MRK coordinates,
\be
\rho _1 = -\frac{z_1 z_2}{1-z_2}\,,\quad \rho _2 = \left(1-z_1\right) z_2\,.
\ee
Similarly, NLLA corresponds to $i_1+i_2=\ell-2$, and the relevant coefficients that are visible at the level of the symbol (imaginary part) may be found in the file \texttt{gTilde.m} attached to \cite{DelDuca:2018hrv}. Note that since both $\cR$ and $\delta_7$ are proportional to $\pi$, beyond one loop we can completely neglect the contribution of the phase in the left-hand side of \eqref{eq:perturbativeExpansion} to the symbol, so that
\be\label{eq:SRtogMRK}
\cS \left(\tfrac{1}{2\pi i}{\cR^{(\ell)}_{h_1,h_2,h_3}}\right)=\sum_{i_1=0}^{\ell-1}\sum_{i_{2}=0}^{\ell-1-i_1} \,\left(\prod_{k=1}^{2}\frac{1}{i_k!}\log^{i_k}\tau_k\right) \,\cS\left({\tilde g}_{h_1,h_2,h_{3}}^{(\ell;i_1,i_2)}\right) \,,\quad \ell\ge 2\,.
\ee
The perturbative coefficients belong to the class of single-valued multiple polylogarithms (SVMPL) \cite{BrownSVMPLs,Brown_Notes,DelDuca:2016lad}, which enjoy the important property that they can be uniquely determined from the knowledge of their holomorphic part, defined as their $\bar z_i\to 0$ limit, also with any divergent $\log\bar z_i$ terms removed. Thus, in order to simplify our comparison even further, we may consider the holomorphic part of \eqref{eq:SRtogMRK}, which for the left-hand side amounts to setting all $\bar z$-dependent factors to one in the limit \eqref{eq:aTozMRK}.

In this manner, we observe perfect agreement between the previously known results for the NMHV heptagon, up to (N)LLA for the $\cR^{(4)}_{+-+}$ ($\cR^{(4)}_{-++}$) helicity configuration, and the corresponding multi-Regge limit of our 4-loop symbol with general kinematic dependence. We view this as strong evidence for the correctness of our result for the latter, as well as of the all-loop dispersion integral \eqref{eq:Rhhhstart}-\eqref{eq:f_hhh}.

Perhaps more importantly, from our calculation we have obtained new predictions for the symbols of the remaining perturbative coefficients in  \eqref{eq:perturbativeExpansion} or \eqref{eq:SRtogMRK}, namely up to N$^3$LLA at 4 loops, which we include as the computer-readable file \texttt{gTilde4L.m} accompanying the version of this paper on the \texttt{arXiv}. These predictions will be useful for determining the central emission block beyond NLO, and may provide significant insight towards its structure to all loops.

\section{Conclusions}
In this paper we presented the computation of the symbol of the
four-loop correction to the NMHV superamplitude of 7 particles in
$\mathcal{N}=4$ super Yang-Mills as the unique combination of weight-8
symbols whose letters are given by a ${\rm Gr}(4,7)$ cluster algebra,
exhibit cluster adjacency in its iterated discontinuities and has a
well-behaved collinear limit. We then analysed the multi-Regge limit of our answer for the amplitude, confirming that it agrees with results derived for the latter up to next-to-leading logarithmic accuracy \cite{DelDuca:2016lad,DelDuca:2018hrv} based on the BFKL approach, and also obtaining new  predictions for an additional two logarithmic orders.

The a priori knowledge of cluster adjacency was key in our computation
in two ways. Firstly it allowed us to construct an ansatz for the
polylogarithmic components of the amplitude with definite, monomial
(final entry)$\otimes$(R-invariant) pairs. Moreover, it restricts the possible next-to-final entries for each of these pairs, drastically reducing the size of the original ansatz.

A peculiar feature of our ansatz for the heptagon amplitude is that it
requires the inclusion of the entire set of 21 R-invariants to
manifestly exhibit cluster adjacency. As a result, integrability of
the symbol is verifiable only on the 6 identities that these
invariants satisfy. This creates a trade-off between two natural ways
of presenting the symbol of the amplitude: one that manifestly
corresponds to a function and one that reveals its cluster adjacent structure.

In this paper we have only exploited the cluster adjacency of
neighbouring symbol letters. However, as noted in
\cite{Drummond:2018dfd}, integrability of the symbols in a sense
``propagates'' the adjacency of adjacent letters to longer words. One
particular example of this phenomenon is the triplets rule which
predicts the combination of letters that come between a mutation pair
separated by one site as the corresponding $\mathcal{X}$
coordinate. It would be interesting to investigate by how much the a
priori implementation of this rule and possible extensions thereof
facilitate the calculation of higher-loop amplitudes using the
bootstrap approach.

It would be very interesting to tackle amplitudes beyond six and seven
particle scattering in $\mathcal{N}=4$ super Yang-Mills where the
symbol alphabet is given by a finite cluster algebra. For instance, the
applicability of cluster adjacency or extended Steinmann relations to individual Feynman integrals
\cite{Drummond:2017ssj,Caron-Huot:2018dsv} strongly suggests that this is a general feature of local quantum
field theories. Furthermore, cluster adjacency has an imprint on the amplitude also in special kinematics, such as the multi-Regge limit we studied, implying relations even between functions of different logarithmic order. Studying more amplitudes in the light of
cluster adjacency may prove useful in developing a more general picture of their analytic structure.

\section*{Acknowledgements}
\"OCG and GP acknowledge the support of the the Galileo Galilei
Institute in Florence in the context of the workshop “Amplitudes in
the LHC era”. JMD, JAF and \"OCG are supported by ERC grant 648630
IQFT.

\bibliographystyle{JHEP-mod}

\bibliography{biblio}

\end{document}